\documentclass{vgtc}                          % final (conference style)
\ifpdf%                                % if we use pdflatex
  \pdfoutput=1\relax                   % create PDFs from pdfLaTeX
  \usepackage{graphicx}                % allow us to embed graphics files
    \usepackage[table]{xcolor}
  \DeclareGraphicsExtensions{.pdf,.png,.jpg,.jpeg} % for pdflatex we expect .pdf, .png, or .jpg files
\else%                                 % else we use pure latex
  \usepackage{graphicx}                % allow us to embed graphics files
  \DeclareGraphicsExtensions{.eps}     % for pure latex we expect eps files
\fi%

\graphicspath{{figures/}{pictures/}{images/}{./}} % where to search for the images

\usepackage{microtype}                 % use micro-typography (slightly more compact, better to read)
\PassOptionsToPackage{warn}{textcomp}  % to address font issues with \textrightarrow
\usepackage{textcomp}                  % use better special symbols
\usepackage{mathptmx}                  % use matching math font
\usepackage{times}                     % we use Times as the main font
\usepackage{cite}                      % needed to automatically sort the references
\usepackage{tabu}                      % only used for the table example
\usepackage{booktabs}                  % only used for the table example
\onlineid{0}

\vgtccategory{Research}

\vgtcinsertpkg

\title{Beyond Predictive Algorithms in Child Welfare}

\author{Erina Seh-Young Moon\thanks{e-mail: erina.moon@mail.utoronto.ca}\\ %
        \scriptsize University of Toronto %
\and Devansh Saxena\thanks{e-mail: devanshsaxena@cmu.edu}\\ %
     \scriptsize Carnegie Mellon University %
\and Tegan Maharaj\thanks{e-mail: tegan.maharaj@utoronto.ca}\\ %
     \scriptsize University of Toronto %
\and Shion Guha\thanks{e-mail:
shion.guha@utoronto.ca}\\ %
    \scriptsize University of Toronto}

\abstract{Caseworkers in the child welfare (CW) sector use predictive decision-making algorithms built on risk assessment (RA) data to guide and support CW decisions. Researchers have highlighted that RAs can contain biased signals which flatten CW case complexities and that the algorithms may benefit from incorporating contextually rich case narratives, i.e. - the \textbf{casenotes} written by caseworkers. To investigate this hypothesized improvement, we quantitatively deconstructed two commonly used RAs from a United States CW agency. We trained classifier models to compare the predictive validity of RAs with and without casenote narratives and applied computational text analysis on casenotes to highlight topics uncovered in the casenotes. Our study finds that common risk metrics used to assess families and build CWS predictive risk models (PRMs) are unable to predict discharge outcomes for children who are not reunified with their birth parent(s). We also find that although casenotes cannot predict discharge outcomes, they contain contextual case signals. Given the lack of predictive validity of RA scores and casenotes, we propose moving beyond quantitative risk assessments for public sector algorithms and towards using contextual sources of information such as narratives to study public sociotechnical systems.} % end of abstract

\CCScatlist{
  \CCScatTwelve{Human-centered computing}{Human computer interaction (HCI)}{Empirical studies in HCI}{};
  \CCScatTwelve{Human-centered computing}{Applied computing}{Computing in government}{}
}

\definecolor{mygray}{gray}{0.25}
\newenvironment{myquote}%
  {\list{}{\leftmargin=0.1in\rightmargin=0.1in}\item[]}%
  {\endlist}

\begin{document}



\firstsection{Introduction}

\maketitle

Predictive risk models (PRMs) have increasingly infiltrated our everyday lives, including within the criminal justice system \cite{angwin, clancy2022}, homelessness \cite{moon24}, employment services \cite{marks22, wieringa20}, the child welfare system \cite{saxena2020human, vaithianathan2017developing}, and education sector \cite{kelly23}, to study and predict the future given the presently available information and generate decisions that carry significant ramifications for individuals. While initially developed with the expectation that such systems can reduce costs and generate objective decision-making \cite{burrell21}, recent research has become increasingly critical of their objectivity and impact. 

The US child welfare (CW) sector is one such sector that has extensively adopted PRMs to predict child maltreatment risk. Given the high stakes for the lives of children and families, such algorithmic systems have been introduced to alleviate caseworkers' high workloads, minimize costs, and arrive at objective, evidence-based decisions with a particular focus on empirical risk measurement and prediction \cite{bosk2018counts, saxena2020human, schwalbe2008strengthening, shlonsky2005next, saxena2021framework2}, i.e., proactively recognizing high-risk cases. However, recent research that examines stakeholder perceptions towards these systems has found that they use potentially biased quantitative inputs related to factors like poverty and unjustly surveil and stigmatize low-income and minority communities when they enter the CW system \cite{stapleton22, eubanks2018automating, tornapart, gerchick23}. There is also increasing awareness in the general scientific community on the relevance of negative results \cite{saxena2023, wilson11, pubbias14} to evaluate and audit the impact of algorithmic systems through multiple methodological lenses. 

In recent years, scholars have increasingly explored how CW narrative data can be used to infer contextual insights around the CW system and be incorporated into PRMs. Saxena et al. applied topic modeling on caseworker casenotes to illuminate invisible patterns of frontline caseworker labor \cite{saxena22} and uncovered that while risk in CW is primarily measured as a function of \textit{risk introduced by clients}, this does not fully account for a clients' holistic circumstances and the dynamic multiplicity/temporality of risk factors underpinning CW cases \cite{saxena2023}. Field et al. \cite{field23} also found training computational models using text data \textbf{with} quantitative, structural administrative data (e.g., demographic information, past public welfare receipt history, criminal history) may mitigate racial disparities in out-of-home placements predictions. With these works in mind, we critically examined the predictive validity of the \textit{conceptualization of risk} in predictive risk models prevalent in the child welfare sector (CWS). We quantitatively compared signals from RA data and CW narratives to ask \textbf{RQ1:} How does risk measured in CW risk assessments affect case outcomes? \textbf{RQ2:} How do case narratives inform child welfare caseworker decisions and case outcomes? and \textbf{RQ3:} How can researchers better support child welfare caseworkers?

We partnered with a child welfare agency in a US Midwestern state and examined casenotes written by caseworkers and CW risk assessment (RA) scores assessed for 277 families discharged from the agency between November 2018, to March 2022. We trained 14 classifier models using support vector machines (SVM) and random forest models, incorporating narrative data from CW caseworker casenotes using topic modeling and RA scores from the Adult Adolescent Parenting Inventor-2.5 (AAPI) and North Carolina Family Assessment Score (NCFAS) (two commonly used RAs which aims to comprehensively measure case risk factors for families \cite{conners06, lee10}). We compared the predictive validity of our trained classifier models to examine how well casenote data and RA scores can predict reunification (discharge) outcomes. We also qualitatively examined topics in our topic model solution to better understand the types of case signals our topic model solution provided to the trained classification models. In summary, our study makes the following contributions.

\begin{table}[]
\centering
\scalebox{0.95}{
\begin{tabular}{l|l}
\hline
\textbf{Model} & \textbf{True Negative} \\
\textbf{(best-performing in overall accuracy)} & \textbf{Rate}\\ \hline
Random Forest using RA scores (AAPI scores) & 0.167 \\
Random Forest using Topic Model Probabilities & 0.357 \\
Constant Predictor & 0.5 \\
Random Predictor & 0.2 \\ \hline
\end{tabular}}
\vspace{0.1cm}
\caption{Classification model metrics for the top performing models: We show how often the top performing models could identify when children were not reunified with their bio-parent(s) given a balanced dataset. The Constant Predictor predicts reunified (`R') and not reunified (`NR') outcomes 50\% of the time each. The random predictor predicts `R' outcomes 80\% of the time and `NR' 20\% of the time. }
\label{tab:introtable}
\vspace*{-0.5cm}
\end{table}

\begin {itemize}
\item We quantitatively deconstruct AAPI and NCFAS scores, which measure child maltreatment risk across 100 metrics, to examine their ability to correctly predict discharge outcomes. We find quantitative support that RAs are largely irrelevant in informing discharge outcomes. We find that while the best performing classifier built using \textit{only} RA scores shows a high accuracy rate of 86.4\%, the model can correctly identify cases that end in non-reunification (`NR') between child and bio-parent(s) only 16.7\% of the time (seen in Table \ref{tab:introtable}). This is worse than a constant predictor or random predictor. 

\item Through computational text analysis, we show that case narratives provide different signals around cases compared to the RA scores. Case narratives can highlight contextual details between individuals involved in a case that can affect the trajectory and outcome of cases. We find that a random forest model built using \textit{only} topic model solution data can correctly identify cases that end in `NR' 35.7\% of the time (see Table \ref{tab:introtable}). This is better than a random predictor but worse than a constant predictor. 

\item We show how casenotes inherently contain limitations in prediction tasks in CW as they present the writer's perspectives. We also show narratives in aggregation can provide the contextual rationale behind CW decisions. 

\item We computationally critique CW algorithms that use RA scores to predict and affect case outcomes. We find support to shift away from predictive algorithms in CW and towards using contextual information such as narratives to study public sociotechnical systems.

\end{itemize}

\section{Ethical Considerations}
We had Institutional Review Board (IRB) approval from our university to use CW casenotes and discharge records for this research. The data we used for this research was bound by the IRB and a research agreement with the CW agency, which only permits us to use the data for research purposes. We are also aware that beyond the IRB guidelines and data-sharing research agreement, the nature of our research is highly sensitive, and the process in which CW data is collected is a morally complicated matter related to privacy, oppression, and surveillance. Considering the pervasiveness of predictive risk models (PRMs) in CW and their biased impact on families, we hope our study can illuminate the limitations of using RA scores to build and guide CW decisions. In this study, to minimize the risk of unintended harm, we considered the principles of beneficence, respect for law, and public interest when determining how and to what extent we should utilize the information from our datasets for our analysis. We anonymized names and numbers from the data to ensure private information does not leak into our predictive models and do not make any raw data public.

\section{Related Work}

%\subsection{Algorithmic Decision-Making in the Public Sector}
\subsection{Algorithmic “Risk” Assessment Tools in the Public Sector}
Public sector agencies in several countries are experiencing increasing pressures from policymakers to use cross-sector public administrative data to improve decision-making in managing and delivering public services by employing algorithmic systems \cite{saxena2021framework2, holten2020shifting, veale2019administration, allhutter2020algorithmic}. Predictive risk assessment algorithms are now being used to make high-stakes decisions about human lives by several public agencies such as the child welfare system, criminal justice system, public education, housing authority, and public benefits programs to decide which neighborhoods require more surveillance, predict the risk of recidivism \cite{veale2018fairness}, tackle school segregation and assign students to school zones \cite{robertson2020if, robertson2021modeling}, determine student performance \cite{williamson2016digital}, and to assess the risk of long-term unemployment for a person \cite{allhutter2020algorithmic, holten2020shifting}. 

As Redden et al. \cite{redden2020datafied} note, one of the critical drivers in this push for algorithmic systems for governance has been the preemptive recognition and mitigation of “risk”, i.e. – identifying clients in the riskiest circumstances and extending services to them. However, many of these algorithms have inadvertently exacerbated racial biases where minority communities are being over-surveilled \cite{veale2018fairness}, schools are becoming more segregated \cite{nytimes_school}, and minority families are being over-investigated for allegations of child maltreatment \cite{allegheny2022}. On the other hand, due to an overemphasis on clients “most in need”, resources from public benefits programs are being directed away from citizens who only require temporary assistance to regain stability in their lives, i.e. – clients who require significantly fewer government resources but are labeled “low risk” and are consequently, screened out of consideration \cite{eubanks2018automating}. As highlighted by prior work, four critical factors make the task of risk prediction especially hard in the public sector – 1) outcomes in the public sector such as the risk of recidivism and risk of child maltreatment are poorly and inconsistently defined \cite{greene2020hidden, barocas2016big, saxena2020human, eubanks2018automating}, and 2) it is not possible to directly observe and assess an outcome such as “risk of future child maltreatment”, therefore, algorithms are trained to predict proxy outcomes such as the likelihood of re-referral or placement in foster care within the next two years \cite{kawakami22, feng22}. Furthermore, 3) a person’s life can stabilize and destabilize several times which makes it hard to assess what may constitute a successful outcome or intervention \cite{holten2020shifting, saxena2021framework2}, and 4) administrative data in the public sector is poorly and inconsistently collected which leads to several algorithmic biases in outcomes that seek to predict client’s behavior and may lead to unfairness in decision-making processes \cite{connelly2016role, saxena2023, gillingham2018decision}. Therefore, there is a need to critically examine how predictive risk models are formulated in the public sector and whether they are achieving better outcomes for citizens.

\subsection{Predictive Risk Models (PRMs) in Child Welfare}

Predictive risk models (PRMs) are deeply embedded in various stages of the CW system \cite{saxena2021framework2}. Initially introduced to mitigate concerns around subjective decision-making by individuals and to arrive at data-driven, evidence-based, consistent decision-making for CW cases \cite{shlonsky2005next, schwalbe2008strengthening, saxena2020human}, PRMs help determine which families should be screened in for further investigation when an allegation of child maltreatment is made, deciding whether to remove a child from their biological parents and predicting future maltreatment risk \cite{gillingham2018decision, eubanks2018automating, saxena2021framework2}. For a more in-depth analysis of child welfare algorithms, we direct the readers to Saxena et al. \cite{saxena2020human} which provides a comprehensive review of predictive risk models in CWS. 

Recent scholarship studying CW algorithms has largely used a participatory approach to evaluate the downstream effects of implementing these systems on different racial or income groups, envision solutions to mitigate algorithmic harms \cite{stapleton22, feng22, gold19}, and qualitatively understand stakeholder perceptions and usages of these computational systems \cite{kawakami22b, saxena2020human, hornby18}. Notably, Stapleton et al. \cite{stapleton22}, and Brown et al. \cite{brown2019toward} highlighted the harms caused by these models finding that CW stakeholders (e.g., caseworkers, families involved with CWS) have concerns about how the models focus on case deficits (i.e., risk) rather than considering case strengths which can inadvertently perpetuate stigmatization and racial biases. Moreover, researchers have found that subjectivity and caseworker discretion are absorbed into PRMs despite their touted objectivity. Inputs into these models often involve factors that require inherent interpretation such as parent cooperativeness, parenting attitudes, or home visits \cite{saxena2021framework2, bosk2018counts} and systematic factors where due to high staff turnover, inexperienced caseworkers measure risk based on impressions \cite{copeland2021s, bosk2018counts, saxena2022chilbw}. In field studies, Cheng et al. \cite{feng22}, Kawakami et al. \cite{kawakami22b}, and Saxena et al. \cite{saxena2021framework2} found that caseworkers actively exercise discretion to mitigate racially biased outcomes from PRMs, gamed the models, or tried to guess how features affected risk scores to better serve their clients. Saxena et al. also \cite{saxena2023} further problematized empirical CW risk prediction. The authors found that PRMs often treat risk as a static construct when in fact, risk in CW cases is dynamic and caused by multiple factors, including the child welfare system itself which faces a shortage of good foster homes and experienced caseworkers who can better support clients. In sum, several qualitative \cite{kawakami22b, saxena2021framework2}, quantitative \cite{feng22, saxena2022chilbw}, and mixed-methods studies \cite{saxena2023} conducted on PRMs in child welfare have questioned the validity of empirical risk prediction itself.

\subsection{Assessing the validity of Child Welfare PRMs}

Coston et al. \cite{coston2023validity} and Raji et al. \cite{raji22} argue it is important to examine the validity of decision-making algorithms to evaluate their functionality. In a related vein, Saxena et al. \cite{saxena2023, saxena2022chilbw} recently raised construct validity concerns around PRMs in child welfare, finding marked divergences between how risk is quantified as a static construct in risk ratings and how they are depicted as a temporally dynamic construct in CW casenotes. These studies raised the need to better understand and account for contextual systematic and procedural factors to measure `risk' in CW. And yet, interestingly, although PRMs mostly focus on predicting child maltreatment `risk', to the best of our knowledge, few studies quantitatively examine their validity to question how `risk' in these models is conceptualized, measured, and utilized to guide decision-making. To date, Vaithianathan et al. \cite{afsthospital} have examined the predictive validity of a PRM by measuring the association between risk scores and future child hospitalization/emergency department visits in the follow-up period. However, because their study focused on probabilistic associations between risk scores and future maltreatment events, their findings could be subject to confounding factors as CWS would likely intervene once a family is deemed high risk. 

In sum, we find that while many child welfare PRMs focus on measuring and predicting child maltreatment risk, there is a gap in the literature around our understanding of how risk is conceptualized in these models from a validity perspective. In this study, we specifically examined the predictive validity of risk assessments (RAs) used to build PRMs by examining how RAs and casenotes inform discharge outcomes. In CW, discharge outcomes signal whether the courts and caseworkers determine a child is safe to be reunified with their parents. Using machine learning techniques, our study asked how risk measured in RAs affects CW discharge outcomes and how casenotes inform CW decisions and outcomes. Through our research questions, we seek to better understand how researchers can better support CW caseworkers.

\section{Research Context} \label{researchcontext}

We partnered with a CW agency in a Midwestern US state. The agency is contracted by the state's Department of Children and Families (DCF) to support foster homes and offers case management services in compliance with all DCF directives. When allegations of child maltreatment are reported and hereafter substantiated by a DCF worker, the case is referred to the CW agency. Following DCF standards, caseworkers conduct initial assessments to determine the family's needs and provide recommendations to the court. Then, after discussions with birth parent representatives, the judge, and the district attorney's office, CW teams at the agency work together to offer services to families. Within the CW agency, various teams provide different services to families, including teams that focus on offering in-home services, preventing sex trafficking, and helping foster youth transition out of foster care.

For this study, we specifically obtained casenotes and a discharge dataset with RA scores from the Family Preservation Services (FPS) team at the CW agency. When a case is assigned to the FPS team, FPS caseworkers assist bio-parents achieve reunification with their children by providing crisis support, parenting classes, and helping improve family functioning \cite{fps}. Following FPS intervention, FPS caseworkers provide evidence to the DA's office and judge to recommend a discharge outcome for the family, such as reunification with their bio-parent, placement into foster care, kinship care, or adoption. Discharge outcomes on families meaningfully signal whether caseworkers and the courts deem a child safe to be reunited with their bio-parent(s) or if they should be placed into other placement forms such as foster care or adoption \cite{ra_info}. 

An integral part of FPS caseworker duties include recording casenotes following stringent documentation standards guided by social work theory \cite{cpsmanual}. These casenotes serve multiple critical purposes, including, providing evidence for the DA's office to recommend reunification, ensuring accountability to caseworkers and the agency, and helping guide the child welfare process by providing a summary of services offered to families \cite{cpsmanual, sidell15}. DCF also requires FPS caseworkers to complete quantitative RAs to assess family risk and safety per the Child Abuse Prevention and Treatment Act \cite{ra_info}. In the CW agency's Midwestern state, caseworkers are required to conduct RAs at intake and discharge (when families enter and before leaving CW services) to consistently collect information on factors that threaten child safety. Based on the RA scores and casenotes recorded, caseworkers then use casenotes and RAs as evidence to justify case discharge recommendations. 

\begin{table*}[h]
\centering
\begin{tabular}{l|l|l|l}
\hline
\textbf{Dataset Name} & \textbf{N} & \textbf{Features Included} & \textbf{Target Variable} \\ \hline
TM & 277 & 19-topic model solution probabilities (19TM) & Discharge outcome \\
AAPI & 145 & Intake \& discharge AAPI scores & Discharge outcome \\
NCFAS & 221 & Intake \& discharge NCFAS scores & Discharge outcome \\
AAPI + NCFAS & 133 & Intake \& discharge AAPI and NCFAS scores & Discharge outcome \\
AAPI + TM & 145 & Intake \& discharge AAPI scores and 19TM & Discharge outcome \\
NCFAS + TM & 221 & Intake \& discharge NCFAS scores and 19TM & Discharge outcome \\
AAPI + NCFAS + TM & 133 & Intake \& discharge AAPI and NCFAS scores and 19TM & Discharge outcome \\ \hline
\end{tabular}
\caption{Descriptions of the six datasets (N here indicates the number of observations in the dataset)}
\label{tab:6datasets}
\end{table*}

\section{Methods}

\subsection{Dataset} \label{sec:dataset}
For our study, we combined two different datasets provided by the Family Preservation Services (FPS) team at our partner child welfare agency, the discharge dataset and the casenote dataset. %Please refer to Section \ref{researchcontext} for detailed information on the role of the FPS team within the agency. 

\textit{Casenote dataset} The casenotes dataset comprised 12,576 casenote records for 471 families that received services from the FPS team from May 1, 2019, to March 31, 2022. Families were not necessarily involved with the agency for this entire period but engaged with the agency within this timeframe. The dataset included all details pertaining to any caseworker interactions and observations of persons involved in a case (e.g., biological parents, children, foster parents) and internal communications between caseworkers. Interactions could be in the form of phone calls, email messages, and in-person or virtual meetings. Each casenote record included a family ID and narrative casenotes.

\textit{Discharge dataset} The discharge dataset comprised 371 discharge records for 339 families discharged from the FPS team from November 2, 2018, to March 24, 2022. Each discharge record included a Family ID, date of admission and discharge from the program, and RA scores for the AAPI \cite{bavolek19}, and NCFAS \cite{kirk04}. These two assessments are two commonly used RAs in CW that fulfill state DCF requirements and are similar to other CW assessments that categorize risk into various levels\footnote{AAPI (Adult Adolescent Parenting Inventory) has been administrated in over 2 million administrations \cite{aapi_info}, and NCFAS has been used in over 1,000 US agencies and in 20 countries \cite{ncfas_info}.}. The AAPI asks parents to rate their parenting attitudes related to child maltreatment from 1 to 10 where scores can be binned into low (8-10), moderate (4-7), and high risk (1-3) \cite{aapiscale}. In contrast, the NCFAS is completed by caseworkers and asks caseworkers to rate multiple domains of family functioning related to a family's environment, parental capabilities, family interactions, family safety, child well-being, and readiness for reunification \cite{kirk15, kirk04}. NCFAS scores range from -3 to +2 and can also include not applicable (`NA') or unknown (`UK') \cite{ncfasscaledef}. Negative scores indicate challenges, 0 signifies the baseline/adequate level, meaning no interventions are needed, and positive scores mean strengths \cite{kirk04}. By design, the AAPI and NCFAS scores are taken twice as a structured means for caseworkers to compare intake and discharge scores and assess how safety and risk factors in cases have been addressed following CW intervention. However, it is not always possible to collect all RA data. Our dataset showed that some intake and discharge RA scores were missing because families refused to take the assessments or caseworkers could not contact the bio-parents as they were hospitalized, incarcerated, missing, or had moved to a different state. Other times, the caseworkers determined the case was a low-needs case and an assessment was unnecessary or if the family had recently been re-referred to the agency and their scores were already on file.

\subsection{Preprocessing the datasets} \label{sec:groupingscores}
To track each family's entire narrative casenote history, we collated all casenote records in the \textit{Casenote dataset} (N=12,576) by each family ID ordered chronologically (oldest to latest). We cleaned and anonymized the casenotes by removing stopwords, punctuation, numbers, and names. After preprocessing the casenotes, we noted that many of the casenotes were lengthy, with collated casenotes averaging 3,299 words, with 53,303 unique words across the casenotes, and there being 38,748 words in the longest casenote. 

In the \textit{Discharge dataset}, each record detailed the discharge outcome of a case which could be either adoption, guardianship/independent living, reunification, and concurrent permanency plan\footnote{This is when caseworkers are working towards reunifying children with birth parents while simultaneously identifying and working towards other placement options for the child such as guardianship with a relative or adoption in case placement with birth parents is unsuccessful \cite{concurrentperm}}. As we wanted to understand how RAs can meaningfully inform whether it is deemed safe for a child to be reunified with their bio-parent, we grouped the discharge outcomes into two: reunified (`R') and not reunified (`NR') \footnote{We categorized adoption, concurrent permanency plan, and guardianship/independent living outcomes as `NR' and reunification outcomes as `R'. We categorized concurrent permanency plan as `NR' as this outcome signaled caseworkers anticipated a possibility that children would not be reunified with their biological parent(s).}. After dichotomizing the discharge outcomes, we noticed more than 80\% of cases ended in `R.'

%\begin{table}[]
%\begin{tabular}{l|ll}
%\cline{1-2}
%\textbf{Metric} & \textbf{Value} &  \\ \cline{1-2}
%Casenotes with more than 500 words & 338 &  \\
%Average number of words per casenote & 3,299 &  \\
%Number of words in longest casenote & 38,748 &  \\
%Number of unique words & 53,303 &  \\ \cline{1-2}
%\end{tabular}
%\caption{Corpus metrics }
%\label{tab:statistics}
%\end{table}

%\subsection{Preprocessing the discharge dataset} 
%\subsubsection{Dataset Preprocessing} 

\textit{Combined dataset} Finally, we combined the two preprocessed datasets by using Family ID as the unique identifier so that we could have one set of casenotes and one discharge outcome for each family ID. Due to how data was recorded at the agency and because some families were continuing to receive CW services, some family IDs that had an associated casenote did not have a discharge outcome and vice versa. In this case, we excluded those family IDs, so that we ended up with a collection of  277 discharge records with associated 277 collated casenotes for 277 families.

\subsection{Topic modeling on narrative casenotes}
\begin{figure*}[!ht]
\centering 
\includegraphics[scale=0.35]{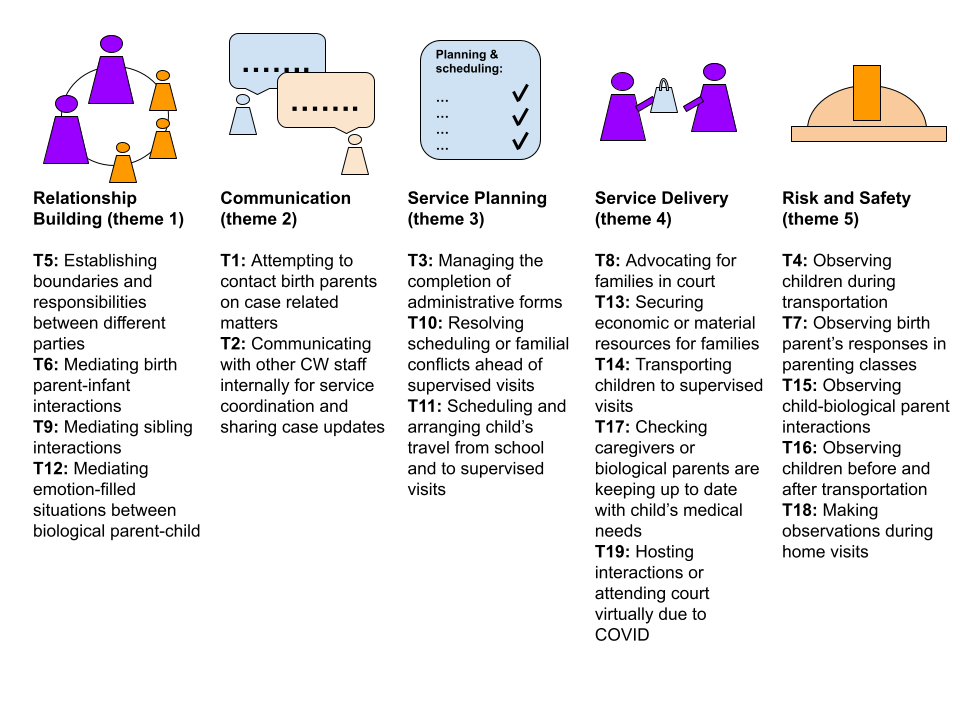}
\vspace{-0.7cm}
\caption{19 topic model solution grouped by 5 high-level themes ('T' stands for topic).}
\label{fig:topic_themes}
\end{figure*}

We adapted the steps taken by Saxena et al. \cite{saxena22} and applied Mallet's implementation of LDA topic modeling \cite{lda_blei} on our collated and preprocessed casenotes for 277 families (from section \ref{sec:groupingscores}). LDA can generate topic distributions over words for topics, produce distributions of topic probabilities for each text document, and scale to dense texts \cite{blei12, antoniak2019narrative, ethno2020, muller2016machine}. Using LDA, we ran a 19-topic model solution, where we computed 19 topic probabilities for each of the 277 collated casenotes. Prior work by Baumer et al. \cite{baumer2017comparing} has argued that conducting grounded thematic methods can help illuminate thematic patterns in text from topic model solutions. Accordingly, to contextually understand topics generated from our topic model solution, the first author of this paper then used an open-coding process to manually inspect casenotes with the highest probabilities associated with each of the 19 topics and keywords associated with the topics to understand and label each topic \cite{braunclark2006}. After labeling each topic, we grouped topics together if they carried similar thematic content to better gauge the high-level themes that emerge from the casenotes (as seen in topic groupings by theme in Fig \ref{fig:topic_themes}).

%Because our study partnered with a CW agency that operates within the same child welfare system as Saxena et al. \cite{saxena22}, the first author compared topics uncovered in the prior study with topics from our topic model solution. The previous study \cite{saxena22} had previously member-checked their interpretations of their topic model solution. Therefore, by comparing topics with theirs, we could determine how valid and realistic our interpretations were by identifying differences and similarities with our topic model solution. 

\subsection{Dataset Creation}
To compare and contrast how different combinations of RA (risk assessment) scores and narratives can inform discharge outcomes, we created six different datasets from the combined dataset outlined in section \ref{sec:groupingscores}. Each dataset contained a combination of intake and discharge RA scores or topic model solution probabilities as features and the discharge outcome for the respective family ID. Table \ref{tab:6datasets} shows the names and data included in the datasets.

\subsection{Classification Models}
We trained classification models using the six different datasets presented in Table \ref{tab:6datasets} to understand how RA metrics can inform CW discharge decisions. We one-hot encoded all intake and discharge AAPI and NCFAS scores in the datasets and then created train and test sets for our datasets using a 7:3 ratio. Training and evaluating risk assessment classification models often carry challenges due to class imbalances common in risk assessment contexts. Class imbalances can cause models to fail to identify patterns from the minority class and instead favor the majority class \cite{smote, japkowicz02, mikel12}. Because our datasets were relatively small and imbalanced (more than 80\% of cases in the datasets ended in `R'), we thus resampled and balanced the training datasets to include 1000 rows of training data where 500 observations ended in `R' and 500 observations ended in `NR' using synthetic minority oversampling techniques (SMOTE) \cite{smote}. SMOTE is useful because its variants can be applied to both continuous and categorical data \cite{smote}, and for each minority class observation, the technique can quickly generate new examples by randomly finding closest neighbors in the feature space \cite{chawla2002}. Next, we reduced the dimensionality of our datasets by applying principal component analysis (PCA) to denoise the datasets \cite{nguyen19}. Finally, using the resampled and dimension-reduced datasets, we trained and compared the performance metrics of random forest models setting a threshold at 0.5 \cite{breiman01} and support vector machines (SVM) using a radial basis function kernel \cite{cortes1995} for each of the datasets. We trained classifiers using random forests and SVMs because they can work with high-dimensional data efficiently and on problems that are non-linear \cite{cohen21, lei17, noble06}. While studies often adjust decision thresholds and validate risk assessment classifiers by inspecting the AUC of the classifier across different thresholds, Lobo et al. \cite{lobo08} discussed by Gerchick et al. \cite{gerchick23}, and Kwegyir-Aggrey \cite{aggrey23} find (1) model comparisons with AUCs obfuscate different error types from different thresholds, (2) determining thresholds based on AUC ignores the inherently normative task of assigning costs to incorrect classifications, and (3) AUCs provide model performance summaries over the receiver operating characteristic (ROC) curve that may not be of relevance. In light of these findings, we evaluated the performance of the classifiers by inspecting the accuracy, false positive rate, false negative rate, and specificity rate of each of the models.  

\vspace{0.2cm}
\section{Results}
In this section, we present our findings from the trained classification models and organized by our research questions. 

\begin{table*}[t]
\centering
\begin{tabular}{l|l|l|l|l|l}
\hline
\multicolumn{1}{c|}{\textbf{Dataset}} & \multicolumn{1}{c|}{\textbf{Model}} & 
\multicolumn{1}{c|}{\textbf{Accuracy}} & \multicolumn{1}{c|}{\textbf{FPR}} & \multicolumn{1}{c|}{\textbf{FNR}} & \multicolumn{1}{c}{\textbf{Specficity}} \\ \hline
\rowcolor{orange}
AAPI &  SVM & 0.864 & 1 & 0 & 0 \\
\rowcolor{orange}
AAPI &  Random Forest & 0.864 & 0.833 & 0.026 & 0.167 \\
\rowcolor{orange}
NCFAS &  SVM & 0.836 & 1 & 0 & 0 \\
\rowcolor{orange}
NCFAS &  Random Forest & 0.806 & 0.909 & 0.054 & 0.091 \\
\rowcolor{orange}
AAPI + NCFAS &  SVM & 0.825 & 1 & 0 & 0 \\
\rowcolor{orange}
AAPI + NCFAS &  Random Forest & 0.825 & 1 & 0 & 0 \\
\rowcolor{cyan}
TM & Random Forest & 0.702 & 0.643 & 0.229 & 0.357 \\
\rowcolor{cyan}
TM & SVM & 0.643 & 0.714 & 0.286 & 0.286 \\
\rowcolor{lightgray}
AAPI + TM & SVM & 0.841 & 1 & 0.026 & 0 \\
\rowcolor{lightgray}
AAPI + TM & Random Forest & 0.841 & 1 & 0.026 & 0 \\
\rowcolor{lightgray}
NCFAS + TM & SVM & 0.806 & 0.909 & 0.054 & 0.091 \\
\rowcolor{lightgray}
NCFAS + TM & Random Forest & 0.806 & 1 & 0.036 & 0 \\
\rowcolor{lightgray}
AAPI + NCFAS + TM & SVM & 0.825 & 1 & 0 & 0 \\
\rowcolor{lightgray}
AAPI + NCFAS + TM & Random Forest & 0.800 & 1 & 0.030 & 0
\end{tabular}
\vspace{0.2cm}
\caption{Performance metrics for trained classifiers using for the seven datasets. The table shows the datasets used to train and test the model (1st column); the type of machine learning algorithm used (2nd column); and the accuracy, false positive, false negative, and specificity rate of the models (last 4 columns). Note that all datasets are class-balanced prior to training.}
\label{tab:classifier}
\end{table*}

 %the percentage of 'R' cases in the dataset (3rd column)
 
\subsection{Risk assessments do not inform discharge outcomes (RQ1)} \label{ra_outcomes}
As seen in the orange rows in Table \ref{tab:classifier}, we found that while classifiers built using only RA scores had high accuracy rates, the models were predominantly predicting children would be reunified with their bio-parent(s) and could not distinguish between cases where children were not reunified with their bio-parent(s). A false positive rate (FPR) in the predictive models indicates how often the models are predicting that a case would end in `R' (reunification) when actually, the child was not reunified (`NR') with their bio-parent(s) in real life\footnote{FPR =FP/(FP+TN)}. As seen in Table \ref{tab:classifier}, SVM models trained using the AAPI or NCFAS or both AAPI+NCFAS dataset all had a FPR equalling 1. This meant that all cases in these models were miscategorized where all `NR' cases (actual outcome) were predicted as `R' cases. The random forest models trained using different combinations of the RA scores, similarly showed a high FPR, suggesting that RA data was not helping predict reunification outcomes regardless of the machine learning algorithm used. A false negative rate (FNR) in the predictive models indicates how often the models are predicting that children will not be reunified with their bio-parent(s) when in fact, they were reunified \footnote{FNR= FN/(FN+TP)}. We noted that models built on RA data made almost no errors or any errors when predicting `R' cases (FNR equaled 0 or nearly 0), meaning that the models were unable to distinguish cases that ended in `R' and `NR' through the RA scores. \textbf{Collectively, we noted classifier models that were trained using either AAPI (parent perspective) or NCFAS scores (caseworker perspectives), or both AAPI and NCFAS scores, did not improve the models' ability to correctly predict discharge outcomes. These results suggested RA scores were not identifying risk factors that determine case outcomes and were providing misleading signals regarding case risk factors.}

\subsection{Case narratives can provide some signals towards discharge outcomes (RQ2)} \label{tm_outcomes}
We also trained classifier models using the probabilities from our trained topic model solution as features to examine whether narrative data can predict discharge outcomes. The blue rows in Table \ref{tab:classifier} show the model metrics that use only topic model data. A specificity rate in the models indicates how often the models are correctly predicting children are not reunified with their bio-parent(s) given the fact they were actually not reunified\footnote{Specificity is calculated by TN/(TN+FP)}. Table \ref{tab:classifier} shows that classifier models that used topic model data had a lower accuracy rate, lower FPR, and higher FNR compared to models built solely on risk assessment data. Compared to models built using RA data, the lower FPR of 64.3\% (random forest model) and 71.4\% (SVM model) indicated that the models were miscategorizing proportionally fewer actual `NR' cases as `R', and the comparatively higher specificity rates showed that the models were doing a better job at predicting `NR' cases. However, we also noted that models built using TM data had a FNR greater than 0, indicating the models were making several `NR' predictions for actual `R' cases. \textbf{These results suggested that topic model solution probabilities may be providing some indicators that can predict discharge outcomes (particularly in identifying and predicting cases that do not end in reunification) compared to using only risk assessment scores.} Our results resonate with Saxena et al. \cite{saxena2022chilbw} who found differences between RA ratings and risks noted in caseworker casenotes, noting caseworkers were recording different facets of the same case in casenotes compared to RAs.

As shown in the gray rows in Table \ref{tab:classifier}, we found that predictive models built using topic model data and RA scores exhibited similar performance metrics to models built using only RA scores. \textbf{Adding risk assessment data to topic model data did not improve the classifiers' ability to correctly predict `NR' outcomes, further suggesting that while our topic model solution probabilities may be providing potentially more useful signals towards case outcomes, these signals are weak.} All in all, our results suggested topic model data may not be appropriate to predict case outcomes but could provide a unique lens into child welfare cases compared to risk assessment scores.

\subsection{Case narratives provide a holistic lens of inquiry towards discharge outcomes (RQ2)} \label{topicmodelsolution}

To further understand the signals provided by the topic probabilities used in the classifier models, we manually examined original casenotes that had the highest probabilities associated with the topics in our topic model solution. \textbf{Manual inspection of casenotes showed that all casenotes were primarily written in a neutral tone that described facts and observations of case details (in fact, caseworkers in the US are trained to write casenotes in a neutral tone \cite{cpsmanual}). At the same time, we noted these casenotes revealed specific case details through the \textit{lens of caseworkers} as they conduct a diverse range of street-level discretionary work.} These findings added support to findings from Section \ref{tm_outcomes} that case narratives were providing contextual information on cases. As the main purpose of this study was to understand how RA scores and topic model solutions can inform case discharge outcomes, we present our topics by high-level themes to highlight how certain topics shared common themes in a manner consistent with prior literature that have performed topic modeling on CW casenotes \cite{saxena22, saxena2023}. The following paragraphs further explain these themes. We also provide paraphrased exemplar sentences related to these themes.

\textbf{Theme 1} was about case details that arise as caseworkers support birth parents to achieve reunification with their children by improving and strengthening relationships with their children, helping them understand their responsibilities, and fostering teamwork between different parties (i.e., foster parents, friends and family of bio-parents). Specifically, this theme revealed case details that emerged as caseworkers helped biological parents and foster parents establish boundaries and responsibilities (topic 5) \cite{bekaert2021family}, mediated interactions between children and birth parents when emotions ran high (topic 12), and observed and aided interactions between bio-parents and children and between siblings to improve family relations and placement stability (topic 6 and 9).

\begin{myquote}
    \small{Topic 5 paraphrased sentence: \textcolor{mygray}{[Bio-parent] has not been able to hold down a job for more than a month. [Caseworker] spoke to parent about how important it is to have a stable job and what can happen if she does not.}}
\end{myquote}

\begin{myquote}
    \small{Topic 6 paraphrased sentence: \textcolor{mygray}{[Baby] was making noises and bouncing on the jumperoo and [bio-parent] sat down in front of the baby. [Caseworker] encouraged [bio-parent] to talk to the baby and explained that babies love receiving attention and love from their caregivers.}}
\end{myquote}

\textbf{Theme 2} highlighted information that arises when caseworkers communicate regularly with bio-parents and with other caseworkers within the agency to make collaborative decisions on cases. Case information emerged as caseworkers regularly attempted to contact bio-parents regarding case updates, schedule parenting classes and supervised visits, and ensure families get the support they need (topic 1) \cite{bekaert2021family}. We found caseworkers also regularly internally shared details on cases with other caseworkers at the agency to make collective case decisions and coordinate services for families (topic 2). 

\begin{myquote}
    \small{Topic 1 paraphrased sentence: \textcolor{mygray}{[Casewoker] tried to get in touch with [bio-parent] but the bio-parent did not answer the phone so caseworker left a voicemail. }}
\end{myquote}

\begin{myquote}
    \small{Topic 2 paraphrased sentence from an email: \textcolor{mygray}{Hello team. Thanks for providing the update. I am going to try to reach the [bio-parent] for their five-day visit and let the team know when we are scheduling the visit.}}
\end{myquote}

\textbf{Theme 3} showed how caseworkers manage and arrange for catered services to be delivered to their clients. Topics under this theme detailed the paperwork caseworkers helped clients fill out and admin consent forms caseworkers needed to collect from bio-parents so that the children could receive required services such as medical services (topic 3) \cite{medicalconsent} and coordinate travel arrangements for children to supervised visits (topic 11). This theme also included details on how caseworkers addressed scheduling, or familial conflicts between biological parents or caregivers to ensure visits with children were taking place (topic 10) \cite{smith2014strengths}. 

\begin{myquote}
    \small{Topic 10 paraphrased sentence: \textcolor{mygray}{[Caseworker] talked to [bio-parent] regarding [child] and visitation. [Bio-parent] said that bio-parent tried to get in touch with [foster parents] about her kids but was not able to get in touch with them.}}
\end{myquote}

\begin{myquote}
    \small{Topic 11 paraphrased sentence: \textcolor{mygray}{[Foster parents] called [caseworker] to determine if there was a visit today. [Caseworker] confirmed there is one today and had discussed this with [bio-parent] about the visit last week regarding visit details.}}
\end{myquote}

\textbf{Theme 4} highlighted the different types of services CW staff were delivering to different families in addition to transportation and legal support. Casenotes revealed how caseworkers helped families obtain economic or material resources by connecting them to employment training centers, finding kitchen appliances and toys (topic 13), and checking in with biological parents or caregivers to ensure they were keeping up with a child's medical schedule (topic 17). Depending on family needs, this theme highlighted how caseworkers also drove children to supervised visits (topic 14), facilitated virtual interactions due to COVID-19 (topic 19), and advocated for families in in-person or virtual court sessions (topics 8 and 19).  

\begin{myquote}
    \small{Topic 13 paraphrased sentence: \textcolor{mygray}{[Caseworker] will help [bio-parent] secure a weight scale. [Caseworker] left a voicemail to [furniture shop] about the delivery of bio-parent’s stove and fridge.}}
\end{myquote}

\begin{myquote}
    \small{Topic 17 paraphrased sentence: \textcolor{mygray}{In today’s visit, the caseworker noted no safety concerns. [Bio-parent] is up to date with all of [child]’s medical appointments.}}
\end{myquote}

\textbf{Theme 5} was about caseworkers noting potential risks and safety concerns in cases and mediating them with protective factors. Caseworkers observed and recorded how children were dressed and acted before, after, and during transporting children to visits (topics 4 and 16), how biological parents responded to course content and questions during parenting classes (topic 7), and how children and birth parents interacted with each other (topic 15) \cite{abuseneglect_guide}. Caseworkers also recorded the safety and suitability of a home for a child by observing who lived in the house and the presence of appliances, furniture, toys, and food at home (topic 18).

\begin{myquote}
    \small{Topic 7 paraphrased sentence: \textcolor{mygray}{[Caseworker] met with [bio-parent] online and completed a chapter of the Active Parenting Curriculum. [Bio-parent] was engaged in the meeting and it was clear bio-parent had read the material before the meeting.}}
\end{myquote}

\begin{myquote}
    \small{Topic 15 paraphrased sentence: \textcolor{mygray}{[Child] begins to rock in the chair and hitting the chair on the wall. [Bio-parent] tells child to stop and child hits the wall harder. [Bio-parent] moves the chair away and the child begins rocking in the chair slowly.}}
\end{myquote}

\section{Discussion}

Abebe et al. \cite{abebe2020roles}’s “Roles for Computing in Social Change” outline four ways computational work can address social issues. Per the researchers, computing as a \textit{diagnostic} can help us understand social problems that manifest in sociotechnical systems and as a \textit{formalizer} to clarify how rule-based models define social issues via the inputs and outputs of the models. Computing can also offer a \textit{rebuttal} to illuminate technical limitations and a \textit{synecdoche} to highlight existing social problems by bringing renewed focus and attention to the issues. Our study takes on these roles and provides a quantitative commentary on existing literature on CW algorithms to affect change in how researchers study and support work in the child welfare domain and more broadly in the public sector. 

\subsection{A quantitative lens into the gaps in risk assessment data in predicting case risk factors (RQ1)} \label{discpt1}

The poor predictive ability of our trained classifiers from section \ref{ra_outcomes} acts as a \textit{diagnostic} \cite{abebe2020roles} of child welfare decision-making algorithms by raising questions about using RA data to build such models \cite{saxena2020human, saxena2021framework2, field23}. The high false positive rate (FPR) in Table \ref{tab:classifier} quantitatively showed that RA data (i.e., using only AAPI, only NCFAS, or AAPI + NCFAS scores) could not help predict discharge outcomes for cases. Most times, the models predicted `R' (reunified) for most cases and nearly all cases that actually ended in `NR' (not reunified) were incorrectly predicted to end in `R.' The inability of the models to correctly identify `NR' cases suggested that the RA scores were not signaling higher risk for these cases. These results were particularly notable because discharge AAPI and NCFAS scores were included in our datasets to train the classifier models, indicating that even discharge scores collected just before a family is discharged from the program cannot provide meaningful signals to predict case outcomes. These findings suggest that the pre-defined risk domains measured in the RAs flatten a case’s complexity \cite{bosk2018counts} and cannot fully capture pertinent risk factors in cases, raising questions on whether we should be using RA data to inform child welfare outcomes. 

The preponderance of the trained classifier models (built with RA data) in predicting ‘R’ outcomes for cases (shown by the FPR which equals or nearly equals 1 in section \ref{ra_outcomes}) find support for studies highlighting how different caseworkers can give very different risk ratings for the same case depending on an individual’s age, experience, stress level, and personal biases \cite{saxena2022chilbw, bosk2018counts, regehr10, saxena2021framework2}. It is possible that the motivations of the biological parents and caseworkers who completed the RAs may have influenced them to give scores that favorably align with their objectives. Most times, biological parents want to be reunified with their children, and the caseworkers are part of the Family Preservation Services (FPS) team, whose main objective is to support families achieve reunification as well \cite{saxena22}. Previous research on caseworker interactions with CW decision-making algorithms has shown that caseworkers will often adjust algorithmic inputs to serve their needs. For example, due to limited good foster homes and to dissuade foster parents from ending placements to children, caseworkers often adjusted algorithm scores so foster parents can receive higher compensation for their services \cite{saxena2021framework2}. Furthermore, as AAPI and NCFAS were designed to measure theoretically derived risk-related domains from child welfare literature, caseworkers are likely working on helping families make changes to familial issues, which are constructs measured in the RAs \cite{kirk04, conners06}. In this case, RA scores may be merely measuring the biological parent’s responses to caseworker interventions \cite{saxena2022chilbw}. Jacobs and Wallach \cite{jacobs20} argue harms related to fairness can arise in computational systems where there exists a mismatch in the intended construct to be measured and its operationalization. Consistent with this argument, our findings question the construct validity of risk prediction models such as the Allegheny Family Screening Tool (AFST) which seeks to predict child maltreatment using proxies such as child welfare records, demographic information, or recent incarceration records \cite{chouldechova2018case, vaithianathan21}, records which in themselves are influenced by potentially biased human discretionary decisions \cite{eubanks2018automating}. 

Prior research examining perceptions surrounding predictive risk models used in child welfare through interviews with affected stakeholders (e.g., family members, caseworkers, and specialists) \cite{brown2019toward, saxena2021framework2, stapleton22, saxena2023algorithmic} have found that stakeholders expressed doubt in the utility of CW algorithmic decision-making systems as they primarily focus on the deficit of cases without accounting for system-level risks. In our study, we find quantitative support for these concerns by showing how RAs can present a reductive representation of complex cases and produce biased scores that are largely irrelevant to discharge outcomes while ignoring structural causes of social issues \cite{keddell15, redden2020datafied}. A \textit{diagnostic} \cite{abebe2020roles} can serve as a starting point for social change by raising awareness of a problem. Through this study, we add quantitative support to existing qualitative concerns surrounding the use of RAs in algorithmic decision-making tools to affect child welfare case outcomes.

\subsection{Shifting away from predictive child welfare predictive analytics by investigating case narratives (RQ2)} \label{discpt2}

Our results from Section \ref{tm_outcomes} show a higher specificity rate for classifiers built using only the TM dataset compared to models built using only RA data. As a \textit{formalizer} \cite{abebe2020roles}, this suggests that compared to RA data, casenotes provide different signals surrounding cases and indicators that can correctly identify ‘NR’ outcomes. A closer dive into the topics in our topic model solution outlined in Section \ref{topicmodelsolution} highlights that unlike the focused risk domains measured by the RAs, the topics presented nuanced details on interactions between multiple agents that can influence the trajectory and outcome of cases. However, as a \textit{rebuttal} \cite{abebe2020roles}, the lower than 50\% specificity rate of the classifiers to correctly predict 'NR' outcomes in Section \ref{tm_outcomes} raises caution against using topic models for predictive tasks. We noted that when we transformed topic model probabilities into continuous features to train classifier models, this step inevitably obfuscated much of the nuanced details recorded in narrative texts. Furthermore, manual inspections of casenotes highly associated with the topics outlined in Section \ref{topicmodelsolution} showed that although caseworkers often wrote detailed casenotes in a neutral tone based on observations and facts, they were still contextualized \textit{caseworker perspectives} that may be biased because they could choose to include or omit specific case details. For example, we found different caseworkers described transporting children to bio-parent visits differently; one could record emotions exhibited by children during transportation, while others focused on the logistics. 

Consistent with findings by Saxena et al. \cite{saxena2023} and Field et al. \cite{field23}, applying textual data for prediction tasks can therefore be problematic as that would mean relying on individual casenotes for prediction. Casenotes are not uniform nor the ground truth. They record socially situated experiences where different participants in the interactions might experience the same events differently due to structural power asymmetries \cite{tornapart, roberts02, dettlaff2011disentangling, dettlaff20, copeland2021s}. Roberts \cite{tornapart} notes that police reports can entail contradictory records on the same CW case where one officer records a house is in fair condition with food while another records the same home is dirty with no food. In a similar vein, caseworkers may omit or underplay the oppression, surveillance, and coercion experienced by biological parents that fundamentally unpin CW interactions \cite{saxena2023}. Furthermore, we noted that casenotes do not include all case details that may be critical to a case due to legal reasons. For example, we found that when a case involved an ongoing criminal investigation, caseworkers recorded that they omitted details following privacy laws \cite{sidell15}.

Despite limitations on applying topic modeling on casenotes for predictive tasks, we identify key opportunities for using computational text analysis on casenotes. Nguyen et al. \cite{howwedo2020} argue that compared to smaller units of texts, texts in the aggregate can often unveil high-level themes. While caseworkers may exhibit different writing styles, our findings suggest that topic modeling on the casenote corpus can be a useful exploratory tool to uncover the dynamic and temporal dimensions in CW cases where family relations are complex and not self-isolated and exogenous organizational, legislative, and policy tensions pose systemic risks \cite{schwalbe2008strengthening, eubanks2018automating}. A common theme that emerges from HCI research that studies how CW workers use CW algorithms is how workers do not only rely on decision-making algorithms because they deem contextual knowledge of cases \cite{ kawakami22, saxena2021framework2, kawakami22b}, and theoretical knowledge of trauma  \cite{saxena2021framework2} with an awareness of organizational or placement constraints \cite{saxena2020child, saxena2021framework2} as critical factors that affect case decisions. Consistent with Saxena at al. \cite{saxena22}, our findings show that child welfare narratives can unveil these contextual factors mentioned above and reveal invisible labor patterns and day-to-day power dynamics between stakeholders of child welfare. Understanding such details can be helpful for developers and researchers to examine how we can better support caseworkers in managing heavy caseloads and improve the quality of services offered to families \cite{kothari2021retention}.

\subsection{Implications for algorithms in the public sector (RQ3)}

Serving as a \textit{synecdoche} \cite{abebe2020roles}, our research brings attention to the long-standing tenet in child welfare, which is: striving to protect children from maltreatment by assessing child maltreatment risk \cite{camasso2013decision} without fully accounting for systemic risks \cite{tornapart}. Through a \textit{diagnostic} \cite{abebe2020roles}, we quantitatively deconstructed RA scores and casenotes, finding that pre-defined risk parameters cannot correctly predict case outcomes, nor are casenotes suitable for predictive tasks in CW. From these findings, we add quantitative support to existing scholarship on decision-making algorithms used in CW \cite{saxena2020human, saxena2021framework2, feng22, samant21}, the criminal justice system \cite{christin17, jacobs20, dressel18}, and unemployment sector \cite{ammitzboll2021street, seidelin22, holten2020shifting} which have been criticized for using proxies that violate construct validity \cite{mothilal2024nonideal}, focusing on risk factors in cases, and leading to outcomes that are biased by disproportionately affecting historically marginalized population groups \cite{eubanks2018automating, angwin}. 

Through the topics uncovered in our topic model solution outlined in section \ref{topicmodelsolution}, we identify opportunities for using computational text analysis to investigate and support workers in public sector sociotechnical systems. The topics outlined in section \ref{topicmodelsolution} showed that aggregated narrative texts could unveil motivations, lived experiences, and work practices of stakeholders – holistic details not provided in RA data \cite{saxena22}. While individual narratives carry inherent limitations wherein they detail potentially biased \textit{perspectives} \cite{das23}, on the flip side, aggregated \textit{perspectives} offer a unique, interpretive lens into contextual lived experiences that form a part of the complexities of sociotechnical systems, including invisible labor patterns, strengths in cases, and daily power asymmetries between stakeholders \cite{saxena22}. 

The nature of public sector work involves frontline workers making discretionary decisions which are often influenced by normative and socially constructed judgments regarding risk and safety \cite{holten2020shifting, bosk2018counts, saxena2022chilbw, holten21ue}. For example, in CW, caseworkers conduct home safety inspections to determine if a home is safe for children based on culturally influenced indicators of safety and neglect (e.g., cleanliness of a home, presence of furniture) \cite{geiger2021foster, ohrc}. As the agency trains caseworkers to include observations of the home that \textit{back up} their safety assessments in casenotes \cite{geiger2021foster, cpsmanual}, we argue narratives can provide a contextual rationale for decisions made. Similar to the child welfare sector, collecting documentation is a key task in the public sector, including job placement centers and the criminal justice system \cite{ammitzboll2021street}. Here, records pertaining to cases record the dynamic negotiations between lawyers and district attorneys in criminal cases \cite{christin17} and memos between job placement caseworkers and unemployed individuals \cite{ammitzboll2021street}. Therefore, extending our study's opportunities for computational text analysis, we propose harnessing NLP techniques on narrative texts when investigating complex public sector sociotechnical systems. Our topics uncovered in section \ref{topicmodelsolution} suggest using narrative texts as a data source can help \textit{humanize} actors in public sector domains and help us critically and holistically understand how computational or technical interventions may or may not support workers \cite{chancellor19}. Furthermore, by moving away from using RA and administrative data points for predictive tasks, we can avoid abstracting ourselves to data points that distill and flatten social and economic problems in the public sector \cite{hcai, chancellor19, aragon2022human}. 

However, here we add a note of caution. Seidelin et al. \cite{seidelin22} find that public sector algorithms are not always used the way they were initially intended, as is the case for RAs used in child welfare \cite{saxena2020human, saxena2021framework2}. Further, when technical interventions are repurposed from one context to the next they may cause unintended consequences \cite{selbst19, saxena2021framework2, saxena2023algorithmic} and automation bias \cite{eubanks2018automating, saxena22}. Just as AAPI and NCFAS scores were reappropriated to predict risk even though they were initially developed as support tools to help caseworkers identify appropriate services and goals for families \cite{kirk04, bavolek19}, it is always possible that new interventions may diverge from their original intentions. Recently, the private sector has introduced products that claim to harness NLP to assist child welfare caseworkers in case management \cite{augintel, augintelvid}. We argue that it is possible that such tools can diverge from their original intentions and be reappropriated amidst political and policy-related factors as has been done in the past \cite{seidelin22, redden2020datafied, saxena2021framework2}. In light of these concerns, we suggest researchers adopt multiple lenses involving both a quantitative and qualitative lens to investigate stakeholder perceptions and the downstream effects of technical interventions.

\section{Conclusion}
In this study, we quantitatively deconstructed RA scores and child welfare narratives to assume the different roles of computing outlined by Abebe et al. \cite{abebe2020roles}. We critique and add support to recent research that questions the construct validity of existing child welfare algorithms built on RA data by using casenotes and RA scores for cases in a US child welfare agency. In our study, we applied topic modeling to casenotes and trained random forest and support vector machines on RA scores and casenotes. We examined the predictive validity of these data sources on CW discharge outcomes. Our results show RAs could not inform discharge outcomes. We also show narrative casenotes contain caseworker perspectives on child welfare cases that are not uniform and may be biased, making casenotes unsuitable for case outcome prediction. We offer supporting evidence to move away from predictive tasks using RA data and instead find support for using contextual data such as casenotes to study sociotechnical systems in child welfare and, more broadly, in the public sector.

\acknowledgments{
This research was supported by the NSERC Discovery Early Career Researcher Grant RGPIN-2022-04570 and the Connaught New Researcher award. Any opinions, findings, conclusions, and recommendations expressed in this material are those of the authors. We are grateful for the anonymous reviewers whose suggestions and comments helped improve the quality of this manuscript.}

\bibliographystyle{abbrv-doi}

\bibliography{template}

\begin{thebibliography}{100}

\bibitem{abebe2020roles}
R.~Abebe, S.~Barocas, J.~Kleinberg, K.~Levy, M.~Raghavan, and D.~G. Robinson.
\newblock Roles for computing in social change.
\newblock In {\em Proceedings of the 2020 Conference on Fairness, Accountability, and Transparency}, pp. 252--260, 2020.

\bibitem{allhutter2020algorithmic}
D.~Allhutter, F.~Cech, F.~Fischer, G.~Grill, and A.~Mager.
\newblock Algorithmic profiling of job seekers in {Austria}: how austerity politics are made effective.
\newblock {\em Front. Big Data 3: 5. doi: 10.3389/fdata}, 2020.

\bibitem{ammitzboll2021street}
A.~Ammitzb{\o}ll~Fl{\"u}gge, T.~Hildebrandt, and N.~H. M{\o}ller.
\newblock Street-level algorithms and ai in bureaucratic decision-making: A caseworker perspective.
\newblock {\em Proceedings of the ACM on Human-Computer Interaction}, 5(CSCW1):1--23, 2021.

\bibitem{angwin}
J.~A. Angwin, J.~Larson, S.~Mattu, and L.~Kirchner.
\newblock {Machine Bias}, May 2016.

\bibitem{antoniak2019narrative}
M.~Antoniak, D.~Mimno, and K.~Levy.
\newblock Narrative paths and negotiation of power in birth stories.
\newblock {\em Proceedings of the ACM on Human-Computer Interaction}, 3(CSCW):1--27, 2019.

\bibitem{aragon2022human}
C.~Aragon, S.~Guha, M.~Kogan, M.~Muller, and G.~Neff.
\newblock {\em Human-Centered Data Science: An Introduction}.
\newblock MIT Press, 2022.

\bibitem{augintel}
Augintel.
\newblock About us: Social impact meets {A.I.}, 2020.

\bibitem{augintelvid}
Augintel.
\newblock Allegheny county {DHS} case study: Unlocking the data in case notes with natural language processing, may 2022.

\bibitem{barocas2016big}
S.~Barocas and A.~D. Selbst.
\newblock Big data's disparate impact.
\newblock {\em Calif. L. Rev.}, 104:671, 2016.

\bibitem{baumer2017comparing}
E.~P. Baumer, D.~Mimno, S.~Guha, E.~Quan, and G.~K. Gay.
\newblock Comparing grounded theory and topic modeling: Extreme divergence or unlikely convergence?
\newblock {\em Journal of the Association for Information Science and Technology}, 68(6):1397--1410, 2017.

\bibitem{bavolek19}
S.~J. Bavolek and R.~G. Keene.
\newblock {AAPI} onine development handbook the adult-adolescent parenting inventory ({AAPI}-2).
\newblock {\em Family Development Resources, Inc}, 2010.

\bibitem{bekaert2021family}
S.~Bekaert, E.~Paavilainen, H.~Scheke, A.~Baldacchino, E.~Jouet, L.~Zablocka-Zytka, B.~Bachi, F.~Bartoli, G.~Carra, R.~Cioni, et~al.
\newblock Family members’ perspectives of child protection services, a metasynthesis of the literature.
\newblock {\em Children and Youth Services Review}, p. 106094, 2021.

\bibitem{blei12}
D.~M. Blei.
\newblock Probabilistic topic models.
\newblock {\em Commun. ACM}, 55(4):77–84, apr 2012. doi: {{%
10\hspace{.1pt}\discretionary{.}{%
}{.}\hspace{.4pt}1145\discretionary{/}{%
}{/}2133806\hspace{.1pt}\discretionary{.}{%
}{.}\hspace{.4pt}2133826}}


\bibitem{lda_blei}
D.~M. Blei, A.~Y. Ng, and M.~I. Jordan.
\newblock Latent dirichlet allocation.
\newblock {\em J. Mach. Learn. Res.}, 3:993–1022, Mar. 2003.

\bibitem{bosk2018counts}
E.~A. Bosk.
\newblock What counts? {Q}uantification, worker judgment, and divergence in child welfare decision making.
\newblock {\em Human Service Organizations: Management, Leadership \& Governance}, 42(2):205--224, 2018.

\bibitem{braunclark2006}
V.~Braun and V.~Clarke.
\newblock Using thematic analysis in psychology.
\newblock {\em Qualitative Research in Psychology}, 3(2):77--101, 2006. doi: {{%
10\hspace{.1pt}\discretionary{.}{%
}{.}\hspace{.4pt}1191\discretionary{/}{%
}{/}1478088706qp063oa}}


\bibitem{breiman01}
L.~Breiman.
\newblock Random forests.
\newblock {\em Machine Learning}, 45(1):5--32, Oct 2001. doi: {{%
10\hspace{.1pt}\discretionary{.}{%
}{.}\hspace{.4pt}1023\discretionary{/}{%
}{/}A\discretionary{:}{%
}{:}1010933404324}}


\bibitem{brown2019toward}
A.~Brown, A.~Chouldechova, E.~Putnam-Hornstein, A.~Tobin, and R.~Vaithianathan.
\newblock Toward algorithmic accountability in public services: A qualitative study of affected community perspectives on algorithmic decision-making in child welfare services.
\newblock In {\em Proceedings of the 2019 CHI Conference on Human Factors in Computing Systems}, p.~41. ACM, 2019.

\bibitem{burrell21}
J.~Burrell and M.~Fourcade.
\newblock The society of algorithms.
\newblock {\em Annual Review of Sociology}, 47(1):213--237, 2021. doi: {{%
10\hspace{.1pt}\discretionary{.}{%
}{.}\hspace{.4pt}1146\discretionary{/}{%
}{/}annurev\discretionary{%
}{-}{-}soc\discretionary{%
}{-}{-}090820\discretionary{%
}{-}{-}020800}}


\bibitem{cohen21}
P.~D. Caie, N.~Dimitriou, and O.~Arandjelović.
\newblock Chapter 8 - precision medicine in digital pathology via image analysis and machine learning.
\newblock In S.~Cohen, ed., {\em Artificial Intelligence and Deep Learning in Pathology}, pp. 149--173. Elsevier, 2021. doi: {{%
10\hspace{.1pt}\discretionary{.}{%
}{.}\hspace{.4pt}1016\discretionary{/}{%
}{/}B978\discretionary{%
}{-}{-}0\discretionary{%
}{-}{-}323\discretionary{%
}{-}{-}67538\discretionary{%
}{-}{-}3\hspace{.1pt}\discretionary{.}{%
}{.}\hspace{.4pt}00008\discretionary{%
}{-}{-}7}}


\bibitem{camasso2013decision}
M.~J. Camasso and R.~Jagannathan.
\newblock Decision making in child protective services: A risky business?
\newblock {\em Risk analysis}, 33(9):1636--1649, 2013.

\bibitem{cpsmanual}
{Capacity Building Center for States}.
\newblock {Child Protective Services}: A guide for caseworkers, 2018.

\bibitem{chancellor19}
S.~Chancellor, E.~P.~S. Baumer, and M.~De~Choudhury.
\newblock Who is the "human" in human-centered machine learning: The case of predicting mental health from social media.
\newblock {\em Proc. ACM Hum.-Comput. Interact.}, 3(CSCW), nov 2019. doi: {{%
10\hspace{.1pt}\discretionary{.}{%
}{.}\hspace{.4pt}1145\discretionary{/}{%
}{/}3359249}}


\bibitem{chawla2002}
N.~V. Chawla, K.~W. Bowyer, L.~O. Hall, and W.~P. Kegelmeyer.
\newblock {SMOTE}: Synthetic minority over-sampling technique.
\newblock {\em Journal of Artificial Intelligence Research}, 16:321--357, jun 2002. doi: {{%
10\hspace{.1pt}\discretionary{.}{%
}{.}\hspace{.4pt}1613\discretionary{/}{%
}{/}jair\hspace{.1pt}\discretionary{.}{%
}{.}\hspace{.4pt}953}}


\bibitem{feng22}
H.-F. Cheng, L.~Stapleton, A.~Kawakami, V.~Sivaraman, Y.~Cheng, D.~Qing, A.~Perer, K.~Holstein, Z.~S. Wu, and H.~Zhu.
\newblock How child welfare workers reduce racial disparities in algorithmic decisions.
\newblock In {\em Proceedings of the 2022 CHI Conference on Human Factors in Computing Systems}, CHI '22. Association for Computing Machinery, New York, NY, USA, 2022. doi: {{%
10\hspace{.1pt}\discretionary{.}{%
}{.}\hspace{.4pt}1145\discretionary{/}{%
}{/}3491102\hspace{.1pt}\discretionary{.}{%
}{.}\hspace{.4pt}3501831}}


\bibitem{fps}
{Child Welfare Information Gateway}.
\newblock {Family Preservation Services}.

\bibitem{abuseneglect_guide}
{Child Welfare Information Gateway}.
\newblock What is child abuse and neglect? {R}ecognizing the signs and symptoms, 2019.

\bibitem{concurrentperm}
{Child Welfare Information Gateway}.
\newblock Concurrent planning for timely permanency for children, 2021.

\bibitem{ra_info}
{Child Welfare Information Gateway}.
\newblock The use of safety and risk assessments in child protection cases, 2022.

\bibitem{chouldechova2018case}
A.~Chouldechova, D.~Benavides-Prado, O.~Fialko, and R.~Vaithianathan.
\newblock A case study of algorithm-assisted decision making in child maltreatment hotline screening decisions.
\newblock In {\em Conference on Fairness, Accountability and Transparency}, pp. 134--148, 2018.

\bibitem{christin17}
A.~Christin.
\newblock Algorithms in practice: Comparing web journalism and criminal justice.
\newblock {\em Big Data \& Society}, 4(2):2053951717718855, 2017. doi: {{%
10\hspace{.1pt}\discretionary{.}{%
}{.}\hspace{.4pt}1177\discretionary{/}{%
}{/}2053951717718855}}


\bibitem{clancy2022}
K.~Clancy, J.~Chudzik, A.~J. Snowden, and S.~Guha.
\newblock Reconciling data-driven crime analysis with human-centered algorithms.
\newblock {\em Cities}, 124:103604, 2022. doi: {{%
10\hspace{.1pt}\discretionary{.}{%
}{.}\hspace{.4pt}1016\discretionary{/}{%
}{/}j\hspace{.1pt}\discretionary{.}{%
}{.}\hspace{.4pt}cities\hspace{.1pt}\discretionary{.}{%
}{.}\hspace{.4pt}2022\hspace{.1pt}\discretionary{.}{%
}{.}\hspace{.4pt}103604}}


\bibitem{ohrc}
O.~H.~R. Commission.
\newblock Under suspicion: Concerns about child welfare, 2017.

\bibitem{connelly2016role}
R.~Connelly, C.~J. Playford, V.~Gayle, and C.~Dibben.
\newblock The role of administrative data in the big data revolution in social science research.
\newblock {\em Social science research}, 59:1--12, 2016.

\bibitem{conners06}
N.~A. Conners, L.~Whiteside-Mansell, D.~Deere, T.~Ledet, and M.~C. Edwards.
\newblock Measuring the potential for child maltreatment: The reliability and validity of the {A}dult {A}dolescent {P}arenting {I}nventory—2.
\newblock {\em Child Abuse \& Neglect}, 30(1):39--53, 2006. doi: {{%
10\hspace{.1pt}\discretionary{.}{%
}{.}\hspace{.4pt}1016\discretionary{/}{%
}{/}j\hspace{.1pt}\discretionary{.}{%
}{.}\hspace{.4pt}chiabu\hspace{.1pt}\discretionary{.}{%
}{.}\hspace{.4pt}2005\hspace{.1pt}\discretionary{.}{%
}{.}\hspace{.4pt}08\hspace{.1pt}\discretionary{.}{%
}{.}\hspace{.4pt}011}}


\bibitem{copeland2021s}
V.~A. Copeland.
\newblock “{I}t's the {O}nly {S}ystem {W}e've {G}ot”: Exploring emergency response decision-making in child welfare.
\newblock {\em Columbia Journal of Race and Law}, 11(3):43--74, 2021.

\bibitem{cortes1995}
C.~Cortes and V.~Vapnik.
\newblock Support-vector networks.
\newblock {\em Machine Learning}, 20(3):273--297, Sep 1995. doi: {{%
10\hspace{.1pt}\discretionary{.}{%
}{.}\hspace{.4pt}1007\discretionary{/}{%
}{/}BF00994018}}


\bibitem{coston2023validity}
A.~Coston, A.~Kawakami, H.~Zhu, K.~Holstein, and H.~Heidari.
\newblock A validity perspective on evaluating the justified use of data-driven decision-making algorithms.
\newblock In {\em 2023 IEEE Conference on Secure and Trustworthy Machine Learning (SaTML)}, pp. 690--704. IEEE, 2023.

\bibitem{das23}
D.~Das, S.~Guha, and B.~Semaan.
\newblock Toward cultural bias evaluation datasets: The case of {B}engali gender, religious, and national identity.
\newblock In S.~Dev, V.~Prabhakaran, D.~Adelani, D.~Hovy, and L.~Benotti, eds., {\em Proceedings of the First Workshop on Cross-Cultural Considerations in NLP (C3NLP)}, pp. 68--83. Association for Computational Linguistics, Dubrovnik, Croatia, May 2023. doi: {{%
10\hspace{.1pt}\discretionary{.}{%
}{.}\hspace{.4pt}18653\discretionary{/}{%
}{/}v1\discretionary{/}{%
}{/}2023\hspace{.1pt}\discretionary{.}{%
}{.}\hspace{.4pt}c3nlp\discretionary{%
}{-}{-}1\hspace{.1pt}\discretionary{.}{%
}{.}\hspace{.4pt}8}}


\bibitem{dettlaff20}
A.~J. Dettlaff and R.~Boyd.
\newblock Racial disproportionality and disparities in the child welfare system: Why do they exist, and what can be done to address them?
\newblock {\em The ANNALS of the American Academy of Political and Social Science}, 692(1):253--274, 2020. doi: {{%
10\hspace{.1pt}\discretionary{.}{%
}{.}\hspace{.4pt}1177\discretionary{/}{%
}{/}0002716220980329}}


\bibitem{dettlaff2011disentangling}
A.~J. Dettlaff, S.~L. Rivaux, D.~J. Baumann, J.~D. Fluke, J.~R. Rycraft, and J.~James.
\newblock Disentangling substantiation: The influence of race, income, and risk on the substantiation decision in child welfare.
\newblock {\em Children and Youth Services Review}, 33(9):1630--1637, 2011.

\bibitem{dressel18}
J.~Dressel and H.~Farid.
\newblock The accuracy, fairness, and limits of predicting recidivism.
\newblock {\em Science Advances}, 4(1):eaao5580, 2018. doi: {{%
10\hspace{.1pt}\discretionary{.}{%
}{.}\hspace{.4pt}1126\discretionary{/}{%
}{/}sciadv\hspace{.1pt}\discretionary{.}{%
}{.}\hspace{.4pt}aao5580}}


\bibitem{aapiscale}
Edx.
\newblock {A}dult-{A}dolescent {P}arenting {I}nventory-2 ({AAPI}-2), 2015.

\bibitem{eubanks2018automating}
V.~Eubanks.
\newblock {\em Automating inequality: How high-tech tools profile, police, and punish the poor}.
\newblock St. Martin's Press, 2018.

\bibitem{aapi_info}
{Family Development Resources}.
\newblock Inventory scoring system for assessing parenting practices.

\bibitem{field23}
A.~Field, A.~Coston, N.~Gandhi, A.~Chouldechova, E.~Putnam-Hornstein, D.~Steier, and Y.~Tsvetkov.
\newblock Examining risks of racial biases in {NLP} tools for child protective services.
\newblock In {\em Proceedings of the 2023 ACM Conference on Fairness, Accountability, and Transparency}, FAccT '23, p. 1479–1492. Association for Computing Machinery, New York, NY, USA, 2023. doi: {{%
10\hspace{.1pt}\discretionary{.}{%
}{.}\hspace{.4pt}1145\discretionary{/}{%
}{/}3593013\hspace{.1pt}\discretionary{.}{%
}{.}\hspace{.4pt}3594094}}


\bibitem{pubbias14}
A.~Franco, N.~Malhotra, and G.~Simonovits.
\newblock Publication bias in the social sciences: Unlocking the file drawer.
\newblock {\em Science}, 345(6203):1502--1505, 2014. doi: {{%
10\hspace{.1pt}\discretionary{.}{%
}{.}\hspace{.4pt}1126\discretionary{/}{%
}{/}science\hspace{.1pt}\discretionary{.}{%
}{.}\hspace{.4pt}1255484}}


\bibitem{mikel12}
M.~Galar, A.~Fernandez, E.~Barrenechea, H.~Bustince, and F.~Herrera.
\newblock A review on ensembles for the class imbalance problem: Bagging-, boosting-, and hybrid-based approaches.
\newblock {\em IEEE Transactions on Systems, Man, and Cybernetics, Part C (Applications and Reviews)}, 42(4):463--484, 2012. doi: {{%
10\hspace{.1pt}\discretionary{.}{%
}{.}\hspace{.4pt}1109\discretionary{/}{%
}{/}TSMCC\hspace{.1pt}\discretionary{.}{%
}{.}\hspace{.4pt}2011\hspace{.1pt}\discretionary{.}{%
}{.}\hspace{.4pt}2161285}}


\bibitem{geiger2021foster}
J.~M. Geiger and L.~Schelbe.
\newblock Foster care placement.
\newblock In {\em The Handbook on Child Welfare Practice}, pp. 219--248. Springer, 2021.

\bibitem{gerchick23}
M.~Gerchick, T.~Jegede, T.~Shah, A.~Gutierrez, S.~Beiers, N.~Shemtov, K.~Xu, A.~Samant, and A.~Horowitz.
\newblock The devil is in the details: Interrogating values embedded in the allegheny family screening tool.
\newblock In {\em Proceedings of the 2023 ACM Conference on Fairness, Accountability, and Transparency}, FAccT '23, p. 1292–1310. Association for Computing Machinery, New York, NY, USA, 2023. doi: {{%
10\hspace{.1pt}\discretionary{.}{%
}{.}\hspace{.4pt}1145\discretionary{/}{%
}{/}3593013\hspace{.1pt}\discretionary{.}{%
}{.}\hspace{.4pt}3594081}}


\bibitem{gillingham2018decision}
P.~Gillingham.
\newblock Decision-making about the adoption of information technology in social welfare agencies: Some key considerations.
\newblock {\em European Journal of Social Work}, 21(4):521--529, 2018.

\bibitem{gold19}
J.~D. Goldhaber-Fiebert and L.~Prince.
\newblock Impact evaluation of a predictive risk modeling tool for {A}llegheny county’s child welfare office.
\newblock Technical report, 2019.

\bibitem{greene2020hidden}
T.~Greene, G.~Shmueli, J.~Fell, C.-F. Lin, M.~L. Shope, and H.-W. Liu.
\newblock The hidden inconsistencies introduced by predictive algorithms in judicial decision making.
\newblock {\em arXiv preprint arXiv:2012.00289}, 2020.

\bibitem{allegheny2022}
S.~Ho and G.~Burke.
\newblock An algorithm that screens for child neglect raises concerns., 2022.

\bibitem{holten21ue}
N.~Holten Holten~M\o{}ller, T.~Rask Rask~Nielsen, and C.~Le~Dantec.
\newblock Work of the unemployed: An inquiry into individuals’ experience of data usage in public services and possibilities for their agency.
\newblock In {\em Designing Interactive Systems Conference 2021}, DIS '21, p. 438–448. Association for Computing Machinery, New York, NY, USA, 2021. doi: {{%
10\hspace{.1pt}\discretionary{.}{%
}{.}\hspace{.4pt}1145\discretionary{/}{%
}{/}3461778\hspace{.1pt}\discretionary{.}{%
}{.}\hspace{.4pt}3462003}}


\bibitem{holten2020shifting}
N.~Holten~M{\o}ller, I.~Shklovski, and T.~T. Hildebrandt.
\newblock Shifting concepts of value: Designing algorithmic decision-support systems for public services.
\newblock In {\em Proceedings of the 11th Nordic Conference on Human-Computer Interaction: Shaping Experiences, Shaping Society}, pp. 1--12, 2020.

\bibitem{hornby18}
{Hornby Zeller Associates INC.}
\newblock Allegheny county predictive risk modeling tool implementation: Process evaluation.
\newblock Technical report, 2018.

\bibitem{jacobs20}
A.~Z. Jacobs, S.~L. Blodgett, S.~Barocas, H.~Daum\'{e}, and H.~Wallach.
\newblock The meaning and measurement of bias: Lessons from natural language processing.
\newblock In {\em Proceedings of the 2020 Conference on Fairness, Accountability, and Transparency}, FAT* '20, p. 706. Association for Computing Machinery, New York, NY, USA, 2020. doi: {{%
10\hspace{.1pt}\discretionary{.}{%
}{.}\hspace{.4pt}1145\discretionary{/}{%
}{/}3351095\hspace{.1pt}\discretionary{.}{%
}{.}\hspace{.4pt}3375671}}


\bibitem{japkowicz02}
N.~Japkowicz and S.~Stephen.
\newblock The class imbalance problem: A systematic study.
\newblock {\em Intelligent Data Analysis}, pp. 429--449, 2002.

\bibitem{kawakami22}
A.~Kawakami, V.~Sivaraman, H.-F. Cheng, L.~Stapleton, Y.~Cheng, D.~Qing, A.~Perer, Z.~S. Wu, H.~Zhu, and K.~Holstein.
\newblock Improving human-{A}{I} partnerships in child welfare: Understanding worker practices, challenges, and desires for algorithmic decision support.
\newblock In {\em Proceedings of the 2022 CHI Conference on Human Factors in Computing Systems}, CHI '22. Association for Computing Machinery, New York, NY, USA, 2022. doi: {{%
10\hspace{.1pt}\discretionary{.}{%
}{.}\hspace{.4pt}1145\discretionary{/}{%
}{/}3491102\hspace{.1pt}\discretionary{.}{%
}{.}\hspace{.4pt}3517439}}


\bibitem{kawakami22b}
A.~Kawakami, V.~Sivaraman, L.~Stapleton, H.-F. Cheng, A.~Perer, Z.~S. Wu, H.~Zhu, and K.~Holstein.
\newblock “{W}hy do {I} care what’s similar?” {P}robing challenges in {AI}-assisted child welfare decision-making through worker-{AI} interface design concepts.
\newblock In {\em Designing Interactive Systems Conference}, DIS '22, p. 454–470. Association for Computing Machinery, New York, NY, USA, 2022. doi: {{%
10\hspace{.1pt}\discretionary{.}{%
}{.}\hspace{.4pt}1145\discretionary{/}{%
}{/}3532106\hspace{.1pt}\discretionary{.}{%
}{.}\hspace{.4pt}3533556}}


\bibitem{keddell15}
E.~Keddell.
\newblock The ethics of predictive risk modelling in the {A}otearoa/{N}ew {Z}ealand child welfare context: Child abuse prevention or neo-liberal tool?
\newblock {\em Critical Social Policy}, 35(1):69--88, 2015. doi: {{%
10\hspace{.1pt}\discretionary{.}{%
}{.}\hspace{.4pt}1177\discretionary{/}{%
}{/}0261018314543224}}


\bibitem{kirk15}
R.~S. Kirk.
\newblock Psychometric properties of the trauma and post-trauma well-being assessment domains of the {N}orth {C}arolina {F}amily {A}ssessment {S}cale for {G}eneral and {R}eunification {S}ervices ({NCFAS G+R}).
\newblock {\em Journal of Public Child Welfare}, 9(5):444--462, 2015. doi: {{%
10\hspace{.1pt}\discretionary{.}{%
}{.}\hspace{.4pt}1080\discretionary{/}{%
}{/}15548732\hspace{.1pt}\discretionary{.}{%
}{.}\hspace{.4pt}2015\hspace{.1pt}\discretionary{.}{%
}{.}\hspace{.4pt}1090364}}


\bibitem{kirk04}
R.~S. Kirk and K.~Reed-Ashcraft.
\newblock {NCFAS North Carolina Family Assessment Scale Research Report}.
\newblock {\em National Family Preservation Network}, pp. 1--22, 2004.

\bibitem{kothari2021retention}
B.~H. Kothari, K.~D. Chandler, A.~Waugh, K.~K. McElvaine, J.~Jaramillo, and S.~Lipscomb.
\newblock Retention of child welfare caseworkers: The role of case severity and workplace resources.
\newblock {\em Children and Youth Services Review}, 126:106039, 2021.

\bibitem{aggrey23}
K.~Kwegyir-Aggrey, M.~Gerchick, M.~Mohan, A.~Horowitz, and S.~Venkatasubramanian.
\newblock The misuse of {AUC}: What high impact risk assessment gets wrong.
\newblock In {\em Proceedings of the 2023 ACM Conference on Fairness, Accountability, and Transparency}, FAccT '23, p. 1570–1583. Association for Computing Machinery, New York, NY, USA, 2023. doi: {{%
10\hspace{.1pt}\discretionary{.}{%
}{.}\hspace{.4pt}1145\discretionary{/}{%
}{/}3593013\hspace{.1pt}\discretionary{.}{%
}{.}\hspace{.4pt}3594100}}


\bibitem{lee10}
B.~R. Lee and M.~A. Lindsey.
\newblock {North Carolina Family Assessment Scale: Measurement Properties for Youth Mental Health Services}.
\newblock {\em Research on Social Work Practice}, 20(2):202--211, 2010. doi: {{%
10\hspace{.1pt}\discretionary{.}{%
}{.}\hspace{.4pt}1177\discretionary{/}{%
}{/}1049731509334180}}


\bibitem{lei17}
Y.~Lei.
\newblock 3 - {I}ndividual intelligent method-based fault diagnosis.
\newblock In Y.~Lei, ed., {\em Intelligent Fault Diagnosis and Remaining Useful Life Prediction of Rotating Machinery}, pp. 67--174. Butterworth-Heinemann, 2017. doi: {{%
10\hspace{.1pt}\discretionary{.}{%
}{.}\hspace{.4pt}1016\discretionary{/}{%
}{/}B978\discretionary{%
}{-}{-}0\discretionary{%
}{-}{-}12\discretionary{%
}{-}{-}811534\discretionary{%
}{-}{-}3\hspace{.1pt}\discretionary{.}{%
}{.}\hspace{.4pt}00003\discretionary{%
}{-}{-}2}}


\bibitem{smote}
G.~Lema{{\^i}}tre, F.~Nogueira, and C.~K. Aridas.
\newblock Imbalanced-learn: A {P}ython toolbox to tackle the curse of imbalanced datasets in machine learning.
\newblock {\em Journal of Machine Learning Research}, 18(17):1--5, 2017.

\bibitem{lobo08}
J.~M. Lobo, A.~Jiménez-Valverde, and R.~Real.
\newblock {AUC}: {A} misleading measure of the performance of predictive distribution models.
\newblock {\em Global Ecology and Biogeography}, 17(2):145–151, 2008. doi: {{%
10\hspace{.1pt}\discretionary{.}{%
}{.}\hspace{.4pt}1111\discretionary{/}{%
}{/}j\hspace{.1pt}\discretionary{.}{%
}{.}\hspace{.4pt}1466\discretionary{%
}{-}{-}8238\hspace{.1pt}\discretionary{.}{%
}{.}\hspace{.4pt}2007\hspace{.1pt}\discretionary{.}{%
}{.}\hspace{.4pt}00358\hspace{.1pt}\discretionary{.}{%
}{.}\hspace{.4pt}x}}


\bibitem{marks22}
P.~Marks.
\newblock Algorithmic hiring needs a human face.
\newblock {\em Commun. ACM}, 65(3):17–19, feb 2022. doi: {{%
10\hspace{.1pt}\discretionary{.}{%
}{.}\hspace{.4pt}1145\discretionary{/}{%
}{/}3510552}}


\bibitem{kelly23}
K.~McConvey, S.~Guha, and A.~Kuzminykh.
\newblock A human-centered review of algorithms in decision-making in higher education.
\newblock In {\em Proceedings of the 2023 CHI Conference on Human Factors in Computing Systems}, CHI '23. Association for Computing Machinery, New York, NY, USA, 2023. doi: {{%
10\hspace{.1pt}\discretionary{.}{%
}{.}\hspace{.4pt}1145\discretionary{/}{%
}{/}3544548\hspace{.1pt}\discretionary{.}{%
}{.}\hspace{.4pt}3580658}}


\bibitem{ncfas_info}
N.~Mickelson, T.~LaLiberte, and K.~Piescher.
\newblock Assessing risk: A comparison of tools for child welfare practice with {I}ndigenous families, may 2017.

\bibitem{moon24}
E.~S.~Y. Moon and S.~Guha.
\newblock A human-centered review of algorithms in homelessness research.
\newblock 2024. doi: {{%
10\hspace{.1pt}\discretionary{.}{%
}{.}\hspace{.4pt}1145\discretionary{/}{%
}{/}3613904\hspace{.1pt}\discretionary{.}{%
}{.}\hspace{.4pt}3642392}}


\bibitem{mothilal2024nonideal}
R.~K. Mothilal, S.~Guha, and S.~I. Ahmed.
\newblock Towards a non-ideal methodological framework for responsible ml, 2024.

\bibitem{muller2016machine}
M.~Muller, S.~Guha, E.~P. Baumer, D.~Mimno, and N.~S. Shami.
\newblock Machine learning and grounded theory method: Convergence, divergence, and combination.
\newblock In {\em Proceedings of the 19th International Conference on Supporting Group Work}, pp. 3--8. ACM, 2016.

\bibitem{ncfasscaledef}
{National Family Preservation Network}.
\newblock {NCFAS North Carolina Family Assessment Scale Scale \& Definitions (v. 2.0)}.
\newblock 2009.

\bibitem{howwedo2020}
D.~Nguyen, M.~Liakata, S.~DeDeo, J.~Eisenstein, D.~Mimno, R.~Tromble, and J.~Winters.
\newblock How we do things with words: Analyzing text as social and cultural data.
\newblock {\em Frontiers in Artificial Intelligence}, 3:62, 2020.

\bibitem{nguyen19}
L.~H. Nguyen and S.~Holmes.
\newblock Ten quick tips for effective dimensionality reduction.
\newblock {\em JPLoS Comput Biol.}, 15(6), june 2019. doi: {{%
10\hspace{.1pt}\discretionary{.}{%
}{.}\hspace{.4pt}1371\discretionary{/}{%
}{/}journal\hspace{.1pt}\discretionary{.}{%
}{.}\hspace{.4pt}pcbi\hspace{.1pt}\discretionary{.}{%
}{.}\hspace{.4pt}1006907}}


\bibitem{noble06}
W.~S. Noble.
\newblock What is a support vector machine?
\newblock {\em Nature Biotechnology}, 24(12):1565--1567, Dec 2006. doi: {{%
10\hspace{.1pt}\discretionary{.}{%
}{.}\hspace{.4pt}1038\discretionary{/}{%
}{/}nbt1206\discretionary{%
}{-}{-}1565}}


\bibitem{ethno2020}
Y.~Ophir, D.~Walter, and E.~R. Marchant.
\newblock {A Collaborative Way of Knowing: Bridging Computational Communication Research and Grounded Theory Ethnography}.
\newblock {\em Journal of Communication}, 70(3):447--472, 2020.

\bibitem{raji22}
I.~D. Raji, I.~E. Kumar, A.~Horowitz, and A.~Selbst.
\newblock The fallacy of {AI} functionality.
\newblock In {\em 2022 ACM Conference on Fairness, Accountability, and Transparency}, FAccT ’22, p. 959–972. Association for Computing Machinery, New York, NY, USA, Jun 2022. doi: {{%
10\hspace{.1pt}\discretionary{.}{%
}{.}\hspace{.4pt}1145\discretionary{/}{%
}{/}3531146\hspace{.1pt}\discretionary{.}{%
}{.}\hspace{.4pt}3533158}}


\bibitem{redden2020datafied}
J.~Redden, L.~Dencik, and H.~Warne.
\newblock Datafied child welfare services: Unpacking politics, economics and power.
\newblock {\em Policy Studies}, 41(5):507--526, 2020.

\bibitem{regehr10}
C.~Regehr, M.~Bogo, A.~Shlonsky, and V.~LeBlanc.
\newblock Confidence and professional judgment in assessing children’s risk of abuse.
\newblock {\em Research on Social Work Practice}, 20(6):621--628, 2010. doi: {{%
10\hspace{.1pt}\discretionary{.}{%
}{.}\hspace{.4pt}1177\discretionary{/}{%
}{/}1049731510368050}}


\bibitem{roberts02}
D.~E. Roberts.
\newblock Racial harm: {D}orothy {R}oberts explains how racism works in the child welfare system.
\newblock {\em Colorlines}, 5(3):19, Fall 2002.

\bibitem{tornapart}
D.~E. Roberts.
\newblock {\em Torn Apart How the Child Welfare System Destroys Black Families--and How Abolition Can Build a Safer World}.
\newblock Basic Books, New York, NY, USA, 2022.

\bibitem{robertson2021modeling}
S.~Robertson, T.~Nguyen, and N.~Salehi.
\newblock Modeling assumptions clash with the real world: Transparency, equity, and community challenges for student assignment algorithms.
\newblock {\em arXiv preprint arXiv:2101.10367}, 2021.

\bibitem{robertson2020if}
S.~Robertson and N.~Salehi.
\newblock What if {I} don't like any of the choices? {T}he limits of preference elicitation for participatory algorithm design.
\newblock {\em arXiv preprint arXiv:2007.06718}, 2020.

\bibitem{samant21}
A.~Samant, A.~Horowitz, S.~Beiers, and K.~Xu.
\newblock Family surveillance by algorithm: The rapidly spreading tools few have heard of, 2021.

\bibitem{saxena2020child}
D.~Saxena, K.~Badillo-Urquiola, P.~Wisniewski, and S.~Guha.
\newblock Child welfare system: Interaction of policy, practice and algorithms.
\newblock In {\em Companion Proceedings of the 2020 ACM International Conference on Supporting Group Work}, pp. 119--122, 2020.

\bibitem{saxena2021framework2}
D.~Saxena, K.~Badillo-Urquiola, P.~Wisniewski, and S.~Guha.
\newblock A framework of high-stakes algorithmic decision-making for the public sector developed through a case study of child-welfare.
\newblock {\em Proceedings of the ACM on Human-Computer Interaction}, 5(CSCW2), 2021.

\bibitem{saxena2020human}
D.~Saxena, K.~Badillo-Urquiola, P.~J. Wisniewski, and S.~Guha.
\newblock A human-centered review of algorithms used within the us child welfare system.
\newblock In {\em Proceedings of the 2020 CHI Conference on Human Factors in Computing Systems}, pp. 1--15, 2020.

\bibitem{saxena2023algorithmic}
D.~Saxena and S.~Guha.
\newblock Algorithmic harms in child welfare: Uncertainties in practice, organization, and street-level decision-making.
\newblock {\em ACM J. Responsib. Comput.}, sep 2023. doi: {{%
10\hspace{.1pt}\discretionary{.}{%
}{.}\hspace{.4pt}1145\discretionary{/}{%
}{/}3616473}}


\bibitem{saxena2023}
D.~Saxena, E.~S.-Y. Moon, A.~Chaurasia, Y.~Guan, and S.~Guha.
\newblock Rethinking "{R}isk" in algorithmic systems through a computational narrative analysis of casenotes in child-welfare.
\newblock In {\em Proceedings of the 2023 CHI Conference on Human Factors in Computing Systems}, pp. 1--19, 2023.

\bibitem{saxena22}
D.~Saxena, S.~Y. Moon, D.~Shehata, and S.~Guha.
\newblock Unpacking invisible work practices, constraints, and latent power relationships in child welfare through casenote analysis.
\newblock CHI '22. Association for Computing Machinery, New York, NY, USA, 2022. doi: {{%
10\hspace{.1pt}\discretionary{.}{%
}{.}\hspace{.4pt}1145\discretionary{/}{%
}{/}3491102\hspace{.1pt}\discretionary{.}{%
}{.}\hspace{.4pt}3517742}}


\bibitem{saxena2022chilbw}
D.~Saxena, C.~Repaci, M.~D. Sage, and S.~Guha.
\newblock How to train a (bad) algorithmic caseworker: A quantitative deconstruction of risk assessments in child welfare.
\newblock In {\em CHI Conference on Human Factors in Computing Systems Extended Abstracts}, pp. 1--7, 2022.

\bibitem{schwalbe2008strengthening}
C.~S. Schwalbe.
\newblock Strengthening the integration of actuarial risk assessment with clinical judgment in an evidence based practice framework.
\newblock {\em Children and Youth Services Review}, 30(12):1458--1464, 2008.

\bibitem{seidelin22}
C.~Seidelin, T.~Moreau, I.~Shklovski, and N.~Holten~M\o{}ller.
\newblock Auditing risk prediction of long-term unemployment.
\newblock {\em Proc. ACM Hum.-Comput. Interact.}, 6(GROUP), jan 2022. doi: {{%
10\hspace{.1pt}\discretionary{.}{%
}{.}\hspace{.4pt}1145\discretionary{/}{%
}{/}3492827}}


\bibitem{selbst19}
A.~D. Selbst, D.~Boyd, S.~A. Friedler, S.~Venkatasubramanian, and J.~Vertesi.
\newblock Fairness and abstraction in sociotechnical systems.
\newblock In {\em Proceedings of the Conference on Fairness, Accountability, and Transparency}, FAT* '19, p. 59–68. Association for Computing Machinery, New York, NY, USA, 2019. doi: {{%
10\hspace{.1pt}\discretionary{.}{%
}{.}\hspace{.4pt}1145\discretionary{/}{%
}{/}3287560\hspace{.1pt}\discretionary{.}{%
}{.}\hspace{.4pt}3287598}}


\bibitem{shlonsky2005next}
A.~Shlonsky and D.~Wagner.
\newblock The next step: Integrating actuarial risk assessment and clinical judgment into an evidence-based practice framework in cps case management.
\newblock {\em Children and youth services review}, 27(4):409--427, 2005.

\bibitem{hcai}
B.~Shneiderman.
\newblock {\em {Human-Centered AI}}.
\newblock Oxford University Press, 01 2022. doi: {{%
10\hspace{.1pt}\discretionary{.}{%
}{.}\hspace{.4pt}1093\discretionary{/}{%
}{/}oso\discretionary{/}{%
}{/}9780192845290\hspace{.1pt}\discretionary{.}{%
}{.}\hspace{.4pt}001\hspace{.1pt}\discretionary{.}{%
}{.}\hspace{.4pt}0001}}


\bibitem{sidell15}
N.~L. Sidell.
\newblock {\em Social Work Documentation}.
\newblock NASW Press, Washington, DC, USA, 2015.

\bibitem{smith2014strengths}
G.~T. Smith, V.~B. Shapiro, R.~W. Sperry, and P.~A. LeBuffe.
\newblock A strengths-based approach to supervised visitation in child welfare.
\newblock {\em Child Care in Practice}, 20(1):98--119, 2014.

\bibitem{stapleton22}
L.~Stapleton, M.~H. Lee, D.~Qing, M.~Wright, A.~Chouldechova, K.~Holstein, Z.~S. Wu, and H.~Zhu.
\newblock Imagining new futures beyond predictive systems in child welfare: {A} qualitative study with impacted stakeholders.
\newblock In {\em FAccT '22: 2022 {ACM} Conference on Fairness, Accountability, and Transparency, Seoul, Republic of Korea, June 21 - 24, 2022}, pp. 1162--1177. {ACM}, 2022. doi: {{%
10\hspace{.1pt}\discretionary{.}{%
}{.}\hspace{.4pt}1145\discretionary{/}{%
}{/}3531146\hspace{.1pt}\discretionary{.}{%
}{.}\hspace{.4pt}3533177}}


\bibitem{medicalconsent}
Z.~Strassburger.
\newblock Medical decision making for youth in the foster care system.
\newblock {\em J. Marshall L. Rev.}, 49(4):1103--1154, 2016.

\bibitem{nytimes_school}
D.~G. N.~Y. Times.
\newblock San {F}rancisco had an ambitious plan to tackle school segregation. {I}t made it worse., April 2019.

\bibitem{vaithianathan21}
R.~Vaithianathan, D.~Benavides-Prado, E.~Dalton, A.~Chouldechova, and E.~Putnam-Hornstein.
\newblock Using a machine learning tool to support high-stakes decisions in child protection.
\newblock {\em AI Magazine}, 42(1):53--60, 2021.

\bibitem{afsthospital}
R.~Vaithianathan, E.~Putnam-Hornstein, A.~Chouldechova, D.~Benavides-Prado, and R.~Berger.
\newblock {Hospital Injury Encounters of Children Identified by a Predictive Risk Model for Screening Child Maltreatment Referrals: Evidence From the Allegheny Family Screening Tool}.
\newblock {\em JAMA pediatrics}, 174(11):e202770, Nov 2020. doi: {{%
10\hspace{.1pt}\discretionary{.}{%
}{.}\hspace{.4pt}1001\discretionary{/}{%
}{/}jamapediatrics\hspace{.1pt}\discretionary{.}{%
}{.}\hspace{.4pt}2020\hspace{.1pt}\discretionary{.}{%
}{.}\hspace{.4pt}2770}}


\bibitem{vaithianathan2017developing}
R.~Vaithianathan, E.~Putnam-Hornstein, N.~Jiang, P.~Nand, and T.~Maloney.
\newblock Developing predictive models to support child maltreatment hotline screening decisions: Allegheny county methodology and implementation.
\newblock {\em Center for Social data Analytics}, 2017.

\bibitem{veale2019administration}
M.~Veale and I.~Brass.
\newblock Administration by algorithm? {P}ublic management meets public sector machine learning.
\newblock (3375391), 2019.

\bibitem{veale2018fairness}
M.~Veale, M.~Van~Kleek, and R.~Binns.
\newblock Fairness and accountability design needs for algorithmic support in high-stakes public sector decision-making.
\newblock In {\em Proceedings of the 2018 chi conference on human factors in computing systems}, pp. 1--14, 2018.

\bibitem{wieringa20}
M.~Wieringa.
\newblock What to account for when accounting for algorithms: A systematic literature review on algorithmic accountability.
\newblock In {\em Proceedings of the 2020 Conference on Fairness, Accountability, and Transparency}, FAT* '20, p. 1–18. Association for Computing Machinery, New York, NY, USA, 2020. doi: {{%
10\hspace{.1pt}\discretionary{.}{%
}{.}\hspace{.4pt}1145\discretionary{/}{%
}{/}3351095\hspace{.1pt}\discretionary{.}{%
}{.}\hspace{.4pt}3372833}}


\bibitem{williamson2016digital}
B.~Williamson.
\newblock Digital education governance: {D}ata visualization, predictive analytics, and ‘real-time’policy instruments.
\newblock {\em Journal of Education Policy}, 31(2):123--141, 2016.

\bibitem{wilson11}
M.~L. Wilson, W.~Mackay, E.~Chi, M.~Bernstein, D.~Russell, and H.~Thimbleby.
\newblock {RepliCHI - CHI Should Be Replicating and Validating Results More: Discuss}.
\newblock In {\em CHI '11 Extended Abstracts on Human Factors in Computing Systems}, CHI EA '11, p. 463–466. Association for Computing Machinery, New York, NY, USA, 2011. doi: {{%
10\hspace{.1pt}\discretionary{.}{%
}{.}\hspace{.4pt}1145\discretionary{/}{%
}{/}1979742\hspace{.1pt}\discretionary{.}{%
}{.}\hspace{.4pt}1979491}}


\end{thebibliography}
\end{document}